\documentstyle[12pt,aasms4]{article}

\slugcomment{{\it Accepted for publication in the Astronomical Journal, August 2000}}

\lefthead{LEE, YOON, \& LEE}
\righthead{THE H$\beta$ INDEX : THE EFFECTS OF HORIZONTAL-BRANCH STARS}

\begin{document}

\title{THE H$\beta$ INDEX AS AN AGE INDICATOR\\
    OF OLD STELLAR SYSTEMS :\\
    THE EFFECTS OF HORIZONTAL-BRANCH STARS}

\author{Hyun-chul Lee, Suk-Jin Yoon, and Young-Wook Lee}
\affil{Center for Space Astrophysics \& Department of Astronomy,\\
    Yonsei University, Shinchon 134, Seoul 120-749, Korea\\
    Electronic mail : (hclee, sjyoon, ywlee)@csa.yonsei.ac.kr}

\begin{abstract}
The strength of the H$\beta$ index has been computed 
for the integrated spectra of model globular clusters 
from the evolutionary population synthesis. 
These models take into account, for the first time, the detailed 
systematic variation of horizontal-branch (HB) morphology 
with age and metallicity. Our models show that the H$\beta$ index 
is significantly affected by the presence of blue HB stars. 
Because of the contribution from blue HB stars, the H$\beta$ 
does not monotonically decrease as metallicity increases at a given age. 
Instead, it reaches a maximum strength when the distribution of HB stars 
is centered around 9500 K, the temperature where the H$\beta$ 
index becomes strongest. Our models indicate that the strength of the 
H$\beta$ index increases as much as 0.75 {\AA} due to the presence of 
blue HB stars. 

The comparison of the recent Keck observations of the globular cluster 
system in the Milky Way Galaxy with those in giant elliptical galaxies, 
NGC 1399 and M87, shows a systematic shift 
in the H$\beta$ against metallicity plane. Our models suggest that 
this systematic difference is understood if the globular cluster systems 
in giant elliptical galaxies are several billion years 
older, in the mean, than the Galactic counterpart. Further observations 
of globular cluster systems in the external galaxies from the large 
ground-based telescopes and space UV facilities will enable to clarify 
whether this difference is indeed due to the age difference or 
other explanations are also possible.
\end{abstract}


\section{INTRODUCTION}

For distant stellar populations, one relies upon the 
integrated colors or spectra to investigate their ages and metallicities 
since individual stars are not resolved. 
In this paper, we will specifically focus on the H$\beta$ index 
that is widely used as an age indicator. It has been highlighted 
because it is (1) the only index that shows the anticorrelation with 
metallicity and (2) sensitive to temperature by reaching a maximum 
strength at around 9500 K (\cite{wor94}). 

Most of the previous works (e.g., \cite{wor94}; \cite{buz95}), 
however, have been done 
on the basis that stars near the main-sequence turnoff 
(MSTO) region are the most dominant sources for the integrated 
strength of H$\beta$. Consequently, without the meticulous consideration 
for stars beyond the red giant branch (RGB), 
they claimed that the strength of H$\beta$ depends on the location 
of the MSTO, which in turn depends on the age at a given metallicity. 
Several investigators, however, have cast some doubt upon the sensitivity of 
the H$\beta$ index given the presence of other warm stars, especially 
blue horizontal-branch (HB) stars (see, e.g., 
\cite{rab82}; \cite{chr83}; \cite{bur84}; 
\cite{buz94}; \cite{def95}; 
\cite{fer95}; \cite{fis95}; 
\cite{bre96}; \cite{lee96}; 
\cite{gre97}; \cite{jor97}).    

On the observational side, it was barely possible to obtain 
low S/N spectra of globular clusters 
in systems outside the Local Group (e.g., \cite{mou87}; 
\cite{mou90}; \cite{huc87}; \cite{bro91}; 
\cite{gri94}). These spectra have only been used for 
kinematic information. With the advent of 10 m-class 
telescopes, however, Kissler-Patig et al. (1998) 
and Cohen, Blakeslee, \& Ryzhov (1998) have successfully obtained relatively 
high S/N spectra that provide reliable line indices 
calibration for globular clusters in NGC 1399 and M87, 
the central giant elliptical galaxies 
in Fornax and Virgo clusters. 
Figure 1 shows the distribution of globular clusters in 
the Milky Way Galaxy and those in 
NGC 1399 in the H$\beta$ versus Mg$_{2}$ diagram. 
At first glance, there seems to be little difference between 
these two systems especially when considering the large uncertainties in 
NGC 1399 globular cluster system. Some careful scrutiny of the 
data, however, indicates that there is a systematic difference 
in the sense that 
the metal-rich clusters in NGC 1399 have higher H$\beta$ compared 
to Galactic counterparts, while the opposite seems to be the case 
for the more metal-poor clusters. In fact, a Kolmogorov-Smirnov (K-S) test 
gives only 2.14\% probability, excluding the two very metal-rich 
NGC 1399 clusters, that the two samples are extracted 
from the same parent population.

This systematic difference has motivated us to explore 
the sensitivity of the integrated H$\beta$ strength 
with detailed population synthesis models that reproduce the 
systematic variation of HB morphology with age and metallicity. 
The HB stars are not unusual components of old stellar systems 
since they are ubiquitously found from the 
color-magnitude diagrams (CMDs) of Galactic 
globular clusters and the stellar evolution theories remarkably 
reproduce them to a certain degree (e.g., Lee, Demarque, \& Zinn 1990, 1994). 
In the following section, our population models without and with considering 
HB stars are presented. Section 3 reinforces the validity of our models 
with HB stars by comparing them to a sample of Galactic globular clusters 
and compares our results with 
the observations of globular cluster systems in the 
Milky Way Galaxy, M31, NGC 1399, and M87. Finally, section 4 discusses 
major implications from our works.

\section{POPULATION MODELS WITHOUT \& WITH HORIZONTAL-BRANCH STARS}

The present models were constructed by using our evolutionary 
population synthesis code, which was developed specifically 
to study the stellar populations in globular clusters. 
For our population models, the Yale Isochrones 
(\cite{dem96}) rescaled for $\alpha$-elements enhancement (\cite{sal93}) 
and the HB evolutionary tracks by Yi, Demarque, \& Kim (1997) 
have been used. For metal-rich ([Fe/H] $>$ $-$ 0.5) 
populations, the helium enrichment parameter, 
$\Delta${\it Y}/$\Delta${\it Z} = 2, is 
assumed (cf. \cite{mae92}; \cite{pag92}). 
The Salpeter (1955) initial mass function (IMF) 
\begin{equation}
d N = C M^{-\chi} dM
\end{equation} 
with $\chi$ = 2.35 is adopted 
for the relative number of stars along the isochrones, but 
the effects of the variations of $\chi$ 
are also tested and described in section 2.2. 
For the conversion from theoretical quantities to observable quantities, 
we have taken the most recently compiled stellar library of Lejeune, 
Cuisinier, \& Buser (1998) in order to cover 
the largest possible ranges in stellar parameters 
such as metallicity, temperature, and gravity. 

The strength of the spectral line indices is calculated 
with either 
\begin{equation}
EW = \Delta\lambda (1 - (F_{\lambda}/F_{C}))
\end{equation}
or
\begin{equation}
Mag = - 2.5 log (F_{\lambda}/F_{C}),
\end{equation}
where $\Delta\lambda$ is an index bandpass and $F_{\lambda}$ and $F_{C}$ 
are the flux in the index bandpass and 
the pseudocontinuum flux in the index bandpass 
(\cite{wfg94}). First, 
for each star (or bunch of stars after the binning) with a 
given metallicity, temperature, and gravity, 
we find the flux values at both pseudocontinuum regions 
for each index from the stellar library. Then we calibrate 
$F_{C}$ at the center of each index 
bandpasses. Second, we calculate either the equivalent width 
(EW; for H$\beta$ and Mg $b$) or the magnitude (Mag; for 
Mg$_{2}$) using Worthey et al.'s (1994) fitting functions. 
Then we solve for $F_{\lambda}$. After $F_{C}$ and $F_{\lambda}$ are all 
summed up for a simple stellar population of a given metallicity 
and age, finally, we compute the integrated strengths of spectral line indices 
using formulae (2) and (3). 
Table 1 lists the Lick/IDS system bandpasses that have been used in 
this work as defined in Burstein et al. (1984). Unless otherwise noted, 
we have used Burstein et al. (1984) definitions in this work to be 
compatible with the data that we are going to fit in section 3. 

We have not taken into account either blue stragglers (BSs) 
or post-HB stars in our models, since it is hard to quantify them 
systematically with metallicity and age because of the poor knowledge 
of their evolution and also due to their scarcity. 
With the inclusion of the BSs, we would expect 
the strength of H$\beta$ to slightly increase. 
We suspect, however, that the post-HB stars would hardly 
affect this index mainly due to their paucity.

\subsection{Population Models without Horizontal-Branch Stars}
                                
Before proceeding to the construction of models with HB stars, 
we have first compared our models without considering HB stars 
with those of Worthey (1994), the most widely used 
models, to make sure that our calculations are consistent 
with previous investigations. For this purpose, 
we have taken the same index bandpasses that Worthey (1994) 
used in his model calculations. It is clear from Figure 2 that 
both models generate very similar results although there are slight 
differences mostly due to the different isochrones employed in the 
two models. It is also worth noting 
that Worthey models which treated HB stars as red clumps 
are similar to our models without HB 
stars as already noted by Rabin (1982) 
(cf. \cite{wor97}, see the caption of their Fig. 6). 
Without the systematic variation of HB morphology 
that we are going to employ in the following subsection, one can 
see from Figure 2 that the H$\beta$ decreases as metallicity and age 
increase simply because the MSTO temperature decreases with these 
variations. 

Recently, there have been increasing number of reports that the 
evolutionary time scale from MS to RGB should be substantially 
decreased if the developments of 
the stellar evolution theories, 
particularly due to the effects of diffusion and the Coulomb correction 
to the equation of state in the isochrones, are employed 
(see, e.g., \cite{cha96}; \cite{sal97}; 
\cite{cas98}). In particular, according to Chaboyer et al. (1996), 
it is suggested that the age of Galactic globular clusters should be reduced 
by 18\% once one considers this improved input physics. Hence, 
a correction was applied in our models to simulate these age 
reduction effects. For example, 15 Gyr isochrones are consequently employed 
to represent 12 Gyr populations. One can 
see this effect on the H$\beta$ index in Figure 3. The 
dashed lines are made when the standard Yale isochrones are used, while 
the solid lines show what we described above after employing 
the correction. It is clear from Figure 3 that the isochrones with 
the improved input physics (solid lines) should result in the decrease 
of H$\beta$ mainly due to the decrease of the MSTO temperature. Hereafter, 
these simulated isochrones are used for further computations even though 
the conclusions drawn below should be relatively insensitive to this 
correction. Furthermore, it should be noted that only relative ages are 
meaningful from this study until the isochrones that incorporate the 
input physics fully consistent with real stars emerge.

\subsection{Population Models with Horizontal-Branch Stars}

In order to include HB stars into our models, first, we need to estimate  
the amounts of mass loss on the RGB. In this study, we have adopted 
Reimers' mass-loss relation (\cite{rei75}). The value of $\eta$, 
the empirical fitting factor in Reimers' mass-loss relation, was 
estimated by matching the tight correlations between [Fe/H] and HB 
morphology type [(B$-$R)/(B+V+R)] (\cite{lee94}) of the inner halo 
Galactic globular clusters (Galactocentric radius $\leq$ 8 kpc, 
filled symbols in Figure 4). 
The value of $\eta$ = 0.65 was determined at the recently favored 
mean age ($\sim$ 12 Gyr, $\Delta$t = 0 Gyr) of them 
in the light of new distance scale as suggested by 
HIPPARCOS (e.g., \cite{fea97}; \cite{gra97}; 
\cite{rei98}; \cite{cha98}) as well as from the recent 
developments of the stellar evolution theories 
(\cite{cha96}; \cite{saw97}; 
\cite{cas99}). The modified 
Gaussian mass distribution (\cite{lee90}; \cite{dem99}), 
\begin{equation}
 \Psi (M) = \Psi_{\circ} [ M - ( \overline{M_{HB}} - \Delta M )]
(M_{RG} - M) exp[- {{( \overline{M_{HB}} - M )^2} \over {2 \sigma^2}}]\\
\end{equation} 
where $\sigma$ is a mass dispersion factor in solar mass, 
$\Psi_{\circ}$ is a normalization factor, and 
$\overline{M_{HB}}$ ($\equiv$ M$_{RG}$ $-$ $\Delta$M) is the mean mass 
of HB stars, was employed with $\sigma$ = 0.02 M$_{\odot}$. 

Figure 4 presents our model isochrones in the plot of 
the HB type as a function of [Fe/H]. Figure 5 vividly shows 
how the isochrones in Figure 4 are generated. 
From the CMDs in the first column of Figure 5, one clearly sees 
the effect of metallicity, the first parameter that governs the 
HB morphology type, in the sense that HB becomes redder as metallicity 
increases. 
One can also see that the age works as the global second parameter that 
characterizes the HB morphology by looking at Figure 5 horizontally, 
in the sense that it becomes bluer with increasing age at a given metallicity 
(see also \cite{lee94}). 

In Figure 6, the variations of H$\beta$ strength as a function of 
metallicity are plotted at given ages based on the model loci of Figure 4. 
It should be noted here that, unlike the models without HB stars 
(dashed lines), the distinct ``wave-like" features (solid lines) 
appear from our models with HB stars. These features 
are clearly understood with the aid of Figure 5. For example, 
refer the five CMDs in the first column of Figure 5 from bottom to top 
and accordingly follow the model line of $\Delta$t = 
0 Gyr in Figure 6 from low to high metallicity. It is evident that (1) 
there is an enhancement in H$\beta$ and it 
becomes a maximum where HB stars are centered around 9500 K 
(($B$ $-$ $V$)$_{o}$ $\sim$ 0), the temperature at 
which the H$\beta$ index becomes strongest, even though 
the temperature of MSTO is getting lower with increasing metallicity, 
(2) after the peak, the H$\beta$ enhancement decreases 
since the mean temperature of HB stars is getting lower, and (3) with only 
cool red HB stars, the strength of H$\beta$ even becomes 
slightly weaker than that from the models without HB stars. 
In addition, it should be noted that the peak of the H$\beta$ 
enhancement moves to higher metallicity as age gets older. 

Therefore, the strength of 
H$\beta$ does not simply decrease with either increasing age or 
increasing metallicity once HB stars are included in the models. 
As Figure 5 clearly demonstrates, the strength 
of H$\beta$ depends not only on the location of the MSTO but also on the 
distribution of HB stars. Now it is clear that the blue HB stars 
around ($B$ $-$ $V$)$_{o}$ $\sim$ 0 are the key contributors 
for the strength of H$\beta$. The differences in H$\beta$ strengths 
between the models with and without HB stars are as much as 0.75 {\AA} 
at the peak. This is in accordance with the empirical 
estimate by Burstein et al. (1984) and Buzzoni, Mantegazza, \& 
Gariboldi (1994), although these authors did not investigate the 
systematic variations we report here. 

It must be noted here that 
Worthey et al.'s (1994) fitting functions are only given till 13,260 K. 
For the stars hotter than 13,260 K, 
we extrapolated their fitting functions for Mg$_{2}$ and Mg $b$. 
For H$\beta$, however, we employed tables by 
Kurucz (1993) that give explicit values for the Balmer 
line strengths at a given metallicity, temperature, and gravity. 
A systematic shift ($\sim$ 6.1 {\AA}) was necessary in order to make 
H$\beta$ values from Kurucz (1993) match those from Worthey et al.'s (1994) 
fitting function at temperature around 13,000 K. 
Note that this treatment takes effect only for blue HB stars from very 
metal-poor and very old populations (the left-side of the peaks 
for model lines of $\Delta$t = + 2 and + 4 Gyr in Figure 6). 
Therefore, the general ``wave-like" features including 
the location of peaks should not be affected by this 
treatment. The fitting functions in this temperature range 
are needed in the near future, however, to better estimate 
the effects from those hot stars on the spectral line indices. 

We also have looked into the dependence of 
strengths of H$\beta$ and Mg$_{2}$ on the initial mass function. 
It is found that the variations of the exponent $\chi$ 
of the initial mass function between 1.35 and 3.35 are negligible, 
in the sense that they roughly correspond to 
the errors estimated from the fitting functions, $\sim$ 0.22 {\AA} 
for H$\beta$ and $\sim$ 0.008 mag for Mg$_{2}$.

\section{COMPARISON WITH OBSERVATIONS}
    
\subsection{TEST OF THE MODELS VIA GALACTIC GLOBULAR CLUSTERS}

The remarkable effects of HB stars on the strength of H$\beta$ 
are elaborated in the previous section. To 
corroborate our results, our models with HB stars are tested 
using observations of Galactic globular clusters, the only
objects that the independent HB morphology types can be evaluated. 
We have chosen a sample within a narrow range of metallicity, 
so that the metallicity effects on the HB morphology can be put aside. 
Then we investigate whether the strengths of H$\beta$ of the sample 
globular clusters are really in connection with their HB morphology. 

In Figure 7, we have fixed the metallicity ([Fe/H] = $-$ 1.535 ; 
dashed lines and $-$ 0.814 ; solid lines, respectively) and calculated 
the strengths of H$\beta$ with varying ages. 
As age increases, the MSTO temperature is getting cooler while 
the HB morphology becomes bluer. It is shown here that 
the strength of H$\beta$ gradually increases as the HB morphology becomes 
bluer surpassing the MSTO temperature variations. 
The thin lines that have actually nothing to do with HB 
types only indicate the level of H$\beta$ strengths without HB stars. 
The sample of Galactic globular clusters within the similar metallicity 
range (data from \cite{bur84}) that already denoted by squares in 
Figure 4 is superposed. It is clear from Figure 7 that (1) 
the inclusion of relevant HB stars is essential in the population models 
for H$\beta$ and (2) our assumption of the age as the global second parameter 
reproduces the observations within the errors. 

What is evident from this test is that the inclusion of 
pertinent HB stars is dominant over the MSTO temperature variations 
for the strengths of H$\beta$ of globular clusters. 
Therefore, one should give great care to HB stars for the
population synthesis models in order to match the observations of
old stellar systems such as globular clusters.

\subsection{COMPARISON WITH GLOBULAR CLUSTER SYSTEMS}

Having confirmed that the detailed modeling of 
HB is crucial 
in the use of H$\beta$ index as an age indicator, 
now we compare our results with observations of globular cluster systems 
in the Milky Way Galaxy, M31, NGC 1399, and M87. It is important to note 
here that all of these observations except for M31 globular 
clusters were carried out at the Keck telescope with 
the identical instrumental configuration, the Low Resolution Imaging 
Spectrograph (LRIS, \cite{oke95}), so that the relative comparisons 
between these cluster systems are not affected by 
systematic and instrumental inhomogeneity. 

First, Figure 8a presents the comparison of 
the globular cluster system in the Milky Way Galaxy 
with that in M31 in the H$\beta$$-$Mg$_{2}$ plane. 
For the Galactic globular cluster system, 
a set of 12 globular clusters that were 
obtained by Cohen, Blakeslee, \& Ryzhov (1998) are used. 
For that in M31, 
15 out of 18 globular clusters in common with 
Burstein et al. (1984) and Huchra et al. (1996) are used. 
We have taken Huchra et al. (1996) data here, since they were 
obtained with higher resolution and high S/N at the 
Multiple Mirror Telescope. Three clusters (G1, G33, and G222; 
G from Sargent et al. (1977)) are excluded because they 
show more than 1 {\AA} differences in H$\beta$ between Burstein et 
al. (1984) and Huchra et al. (1996) observations. 
The errors given in the diagram are the average from the M31 sample. 
It is found from Figure 8a that both globular cluster systems are 
fundamentally not quite different, in terms of 
H$\beta$ strengths as a function of metallicity. They are 
reasonably well traced by our models of $\Delta$t = 0 Gyr, i.e., 
t = 12 Gyr. 

Next, in Figure 8b, the globular cluster system in NGC 1399 is 
compared with that in the the Milky Way Galaxy. 
Despite the still large observational uncertainties 
in NGC 1399 globular cluster system 
obtained by Kissler-Patig et al. (1998), 
the systematic shift that we noticed in section 1 
between these two systems is now understood by the age difference. 
It is inferred from Figure 8b that 
the NGC 1399 globular cluster system is perhaps systematically older, 
in the mean, than the Galactic counterpart by about 4 Gyr. 

In the following diagram, we compare M87 globular cluster system 
with that in the the Milky Way Galaxy. 
For M87 globular clusters, H$\beta$$-$Mg $b$ plane is used 
since Cohen, Blakeslee, \& Ryzhov (1998) preferred narrower 
Mg $b$ index to broader Mg$_{2}$ index due to their observational 
condition. Among 150 sample from Cohen, Blakeslee, \& Ryzhov 
(1998), we have taken 35 globular clusters that have 
relatively higher S/N (QSNR $>$ 50). It appears from Figure 8c that 
the age difference of $\Delta$t = + 3.5 best reproduces the M87 
globular cluster system in the H$\beta$$-$Mg $b$ plane, 
although better data are clearly needed to confirm this. 

We do not claim 
here that all of the globular clusters in giant elliptical 
galaxies are coeval and older than Galactic counterparts. 
In fact, there is now a growing body of evidence that suggests 
age variations among globular clusters in the Milky Way 
(e.g., \cite{lee94}; \cite{sar97}). In this respect, 
we suspect that there would be a similar 
age variations among globular clusters in giant elliptical galaxies. 
However, the quality of the present data, even taken with the 10 m-class 
telescope, is still far from attempting such detailed analysis.

\section{DISCUSSION}

We have found in this work that (1) the integrated H$\beta$ 
strengths of old stellar systems, such as globular clusters, are 
significantly affected by HB stars, and (2) 
the present observational data for the globular cluster systems 
are best understood in terms of systematic age differences 
among them, in the sense that globular clusters in more massive galaxies 
are older. If our age estimation is confirmed to be correct, 
this would indicate that the star formation in denser environments 
has proceeded much more rapidly and efficiently, so that the 
initial epoch of star formation in more massive (and denser) systems 
occurred several billion years earlier than that of the Milky Way 
(see also Lee 1992). 
This is also consistent with the view that a substantial 
population of the massive early-type galaxies formed earlier 
at very high redshift 
(e.g., \cite{lar90}; \cite{mao90}; 
\cite{ben96}; \cite{har98}; \cite{sta98}). 

Further observations of globular cluster systems 
in nearby galaxies from the large ground-based telescopes are 
imperative to manifest this possible systematic age difference. 
In addition, satellite UV photometry could also provide a test 
for the validity of our results presented in 
this paper, because it is expected that older globular clusters 
would reveal the stronger UV fluxes and 
the bluer UV colors as their blue HB stars should be hotter than 
those in the Galactic counterparts at given metallicity 
(see Figure 1 of Park \& Lee (1997)). 
These new observations, together with the detailed population models 
presented here, will undoubtedly help to clarify our 
understanding of the formation epoch of galaxies.

\acknowledgments

Support for this work was provided by the Creative Research Initiatives 
Program of the Korean Ministry of Science and Technology.

\clearpage

\begin{figure}
 \caption{The comparison of Galactic globular clusters (open 
circles; data from Cohen, Blakeslee, \& Ryzhov 1998) 
with those in NGC 1399 (filled circles; data from 
Kissler-Patig et al. 1998) in the H$\beta$ vs. Mg$_{2}$ diagram. 
The observational errors are displayed for clusters in NGC 1399.}
\end{figure}

\begin{figure}
 \caption{The H$\beta$ strengths from our models 
without HB stars (solid lines) are compared to those from 
Worthey (1994) models (dashed lines) as a function of metallicity (Mg$_{2}$) 
for ages of 8, 12, and 17 Gyr. Note that both models generate 
very similar results.}
\end{figure}

\begin{figure}
 \caption{Similar to Fig. 2, but here the standard Yale isochrones 
(dashed lines) are 
systematically shifted in order to simulate 
the effects of recent improvement in input physics 
in the stellar models (see text). $\Delta$t = 0 Gyr 
corresponds to 12 Gyr populations and 
$-$ 4 Gyr and + 4 Gyr indicate 4 Gyr younger 
and 4 Gyr older populations, respectively.}
\end{figure}

\begin{figure}
 \caption{The value of $\eta$ = 0.65, the empirical fitting 
factor in Reimers' mass-loss relation was   
estimated by matching the tight correlations between [Fe/H] and 
HB morphology type of the inner halo Galactic globular clusters 
(Galactocentric radius $\leq$ 8 kpc, filled symbols)   
at the currently favored 
mean age ($\sim$ 12 Gyr, $\Delta$t = 0 Gyr, solid line) of them. 
The open symbols are for the outer halo Galactic globular clusters. 
The dotted line is the predicted model 
relationship for 4 Gyr younger population, while the short-dashed 
and long-dashed lines are those for 2 and 4 Gyr older populations, 
respectively, than the inner halo Galactic globular clusters. 
The symbols denoted by squares are the selected Galactic globular clusters 
from Burstein et al. (1984) that are compared with models in Figure 7. 
Data are from Lee, Demarque, \& Zinn (1994).}
\end{figure}

\begin{figure}
 \caption{The selected synthetic CMDs at three different 
ages defined in Figure 4 are presented here. At each given age, 
five CMDs are shown according to decreasing metallicity. The 
corresponding isochrones from MS to RGB are displayed by 
solid lines in order to indicate the variations of the MSTO. 
Both the HB morphology type 
[(B$-$R)/(B+V+R)] and the strength of H$\beta$ 
are given in each set of CMDs. 
RR Lyrae stars are denoted by crosses. 
Note that the strength of H$\beta$ depends not only on the location 
of the MSTO but also on the distribution of HB stars.}
\end{figure}

\begin{figure}
 \caption{The effects of HB stars on the strength of H$\beta$ 
as predicted from our models. The strength of H$\beta$ is plotted 
against [Fe/H] (top), Mg$_{2}$ (middle), and Mg {\it b} 
(bottom) under four different relative ages ($\Delta$t = $-$ 4, 
0, + 2, and + 4 Gyr, respectively). The dashed 
lines are models without HB stars, while the solid lines 
are those with HB stars based on the model loci of Figure 4 
(see text).}
\end{figure}

\begin{figure}
 \caption{The variations of H$\beta$ strength with HB morphology 
at given metallicity. In the model calculations (thick lines), 
the age is varied 
from young (left) to old (right) to generate various types of 
HB morphology. The thin lines indicate the level of 
H$\beta$ strengths without HB stars. 
A sample of Galactic globular clusters within the similar 
metallicity range 
is superposed here (data from Burstein et al. 1984). 
Two clusters (NGC 6171 and M13 ; filled and open triangles, 
respectively) 
also obtained at the Keck telescope by Cohen, Blakeslee, \& 
Ryzhov (1998) are connected to Burstein et al. (1984) data by straight 
lines. Note that our models with the systematic variation of HB 
morphology trace the 
observational data reasonably well.}
\end{figure}

\begin{figure}
 \caption{The globular cluster system in the Milky Way Galaxy 
is compared with those in M31 (8a) and NGC 1399 (8b) 
in H$\beta$ vs. Mg$_{2}$ diagrams, and that in M87 (8c) in 
H$\beta$ vs. Mg {\it b} diagram. 
All of the observations except for M31 globular clusters were made from the 
Keck telescope using the identical 
instrumental configuration (see text). Our models with HB stars are overlaid. 
Note that the overall distributions of the globular cluster systems 
in NGC 1399 and M87 are 
different from those in the Milky Way Galaxy and M31, indicating that 
the giant ellipticals may contain globular clusters that are 
several billion years older, in the mean, than those in the Milky Way.}
\end{figure}

\clearpage
 
\begin{deluxetable}{crrrrrrrrrrr}
\footnotesize
\tablecaption{Definitions For Spectral Line Indices \label{tbl-1}}
\tablewidth{0pt}
\tablehead{
\colhead{Index} & \colhead{Blue Continuum ({\AA})}   & 
\colhead{Feature Bandpass ({\AA})}  & \colhead{Red Continuum ({\AA})} & 
\colhead{Type}
} 
\startdata
H$\beta$   &4829.50$-$4848.25 &4849.50$-$4877.00 &4878.25$-$4892.00 &EW \nl
Mg$_{2}$   &4897.00$-$4958.25 &5156.00$-$5197.25 &5303.00$-$5366.75 &Mag \nl
Mg {\it b} &5144.50$-$5162.00 &5162.00$-$5193.25 &5193.25$-$5207.00 &EW \nl

\enddata

\end{deluxetable}

\end{document}